\documentclass[aps,pra,preprint,showpacs]{revtex4-1}

\usepackage{amsmath}
\usepackage{amssymb}
\usepackage{graphicx}
\usepackage{color}

\begin{document}

\title{Photoinduced quantum spin and valley Hall effects and orbital
magnetization in monolayer MoS$_{2}$}
\author{M. Tahir\footnote{m.tahir06@alumni.imperial.ac.uk}, A. Manchon, and
U. Schwingenschl\"{o}gl\footnote{udo.schwingenschlogl@kaust.edu.sa,+966(0)544700080}}
\affiliation{PSE Division, KAUST, Thuwal 23955-6900, Kingdom of Saudi Arabia}

\begin{abstract}
We theoretically demonstrate that 100\% valley-polarized transport in
monolayers of MoS$_{2}$ and other group-VI dichalcogenides can be obtained
using off-resonant circularly polarized light. By tuning the intensity of
the off-resonant light the intrinsic band gap in one valley is reduced,
while it is enhanced in the other valley, enabling single valley quantum
transport. As a consequence, we predict (i) enhancement of the longitudinal
electrical conductivity, accompanied by an increase in the spin-polarization
of the flowing electrons, (ii) enhancement of the intrinsic
spin Hall effect, together with a reduction of the intrinsic
valley Hall effect, and (iii) enhancement of the orbital magnetic moment and
orbital magnetization. These mechanisms provide appealing opportunities to
the design of nanoelectronics based on dichalcogenides.
\end{abstract}

\maketitle

\affiliation{PSE Division, KAUST, Thuwal 23955-6900, Kingdom of Saudi Arabia}

\affiliation{PSE Division, KAUST, Thuwal 23955-6900, Kingdom of Saudi Arabia}

Monolayers of the transition metal dichalcogenides
$MX_{2}$ ($M=$ Mo, W; $X=$ S, Se) are emerging as promising materials for a
wide variety of applications in nanoelectronics, due to their exceptional
band structures \cite{1}. In particular, the exfoliation of MoS$_{2}$
monolayers has attracted significant interest since the realization of
field-effect-transistors with high on-off ratio \cite{2}. Monolayers of
$MX_{2}$ can be regarded as semiconductor analogs of graphene \cite{3,4},
resulting in similar phenomena such Fas spin and valley Hall effects \cite%
{5,6}. Indeed, MoS$_{2}$ has a honeycomb lattice with an intrinsic direct
band gap of 1 to 2 eV, which is in the visible range. The band-edge is
located at the energy degenerate valleys (corners of the hexagonal Brillouin
zone) \cite{7,8}.

Thanks to its direct band gap, MoS$_{2}$ is suitable for optical
manipulations and opens access to many optoelectronic applications \cite%
{7,8,9}. It has been predicted that both valley polarization and valley
coherence can be achieved by optical pumping with circularly and linearly
polarized light \cite{5,6,10}. First experimental realizations have been
reported for MoS$_{2}$ and WSe$_{2}$ \cite{11,12,13,14}, suggesting that
monolayers of $MX_{2}$ could be used for integrated valleytronic devices.
Experiments have shown 30\% to 50\% valley polarization with circularly
polarized light in the resonance regime \cite{10,11,12}. Recent works on the
optoelectronic properties of MoS$_{2}$ indicate that the photoresponse of
externally biased phototransistors is driven by conductivity alteration upon
illumination \cite{15,16,17}. A photovoltaic effect has been reported for MoS%
$_{2}$ devices in contact with metallic electrodes that generate large
Schottky barriers \cite{18,19}. In addition, ultrasensitive phototransistors
with improved mobility have been demonstrated in Ref.\ \cite{20}. These
devices show a photoresponsitivity in the 400 to 680 nm range with a maximum
of 880 A/W at a wavelength of 561 nm.

In contrast with the on-resonant optical induction used till now, we propose
in this letter a scheme to employ \emph{off-resonant} light to influence the
band structure and corresponding transport properties of $MX_{2}$
monolayers, enabling 100\% valley polarization. An important motivation is
the development of new experimental probes \cite{21} that make it possible
to access non-equilibrium effects, where time-periodic perturbation due to
circularly polarized light represents a rich and versatile resource for
creating a band gap \cite{22}. An analog has been realized experimentally in
a photonic system \cite{23,24}. We show that the band structure of MoS$_{2}$
is strongly modified by the off-resonant light, resulting in a
valley-dependent tuning of the band gap, which is not achievable by
on-resonant light. We demonstrate analytically that this valley-dependent
band gap results in the emergence of a quantum spin Hall effect, the
reduction of the quantum valley Hall effect, and enhancement of the orbital
magnetic moment and orbital magnetization.

Off-resonant light cannot generate real photon
absorption/emission due to energy conservation, whereas off-resonant light
can affect the electron system by second order virtual photon processes (a
photon is first absorbed/emitted and then emitted/absorbed). When averaged
over time, these processes result in an effective \emph{static} alteration
of the band gap of the system. Therefore, using circularly polarized light,
it is possible to distinctively tune the band gap at the $K$ and $K'$ valleys.
Off-resonance light-induced gap opening has been predicted for graphene and
the surface states of topological insulators \cite{22} as well as for silicene
\cite{new1}, and has been confirmed experimentally \cite{21}.
These studies show that off-resonant light enables quantum phase transitions
in two-dimensional systems.

The charge carriers in MoS$_{2}$ obey a two-dimensional Dirac-like
Hamiltonian \cite{5,6} with large intrinsic direct band gap and strong
spin-orbit-coupling (SOC). We model MoS$_{2}$ by an effective Hamiltonian in
the $xy$-plane in the presence of circularly polarized light, 
\begin{equation}
\hat{H}^{\eta ,s}(t)=v(\eta \hat{\sigma}_{x}\hat{\Pi}_{x}(t)+\hat{\sigma}_{y}%
\hat{\Pi}_{y}(t))+\Delta \hat{\sigma}_{z}-s\eta \lambda \hat{\sigma}%
_{z}+s\eta \lambda \text{.}  \label{1}
\end{equation}%
Here $\eta =\pm 1$ represents the valleys $K$ and $K^{\prime }$,
respectively, $\Delta $ is the mass term that breaks the inversion symmetry,
($\hat{\sigma}_{x}$, $\hat{\sigma}_{y}$, $\hat{\sigma}_{z}$) is the vector
of Pauli matrices (applies to both the valence and conduction bands), $%
\lambda $ is the SOC energy, $s=\pm 1$ represents the up and down spins,
respectively, and $v$ denotes the Fermi velocity of the Dirac fermions. In
our notation, the spin-quantization axis is chosen along the $z$-direction.
We use the gauge in the two-dimensional canonical momentum $\hat{\mathbf{\Pi 
}}(t)=\hat{\mathbf{P}}-e\mathbf{A}(t)$, with the vector potential%
\begin{equation}
\mathbf{A}(t)=(\pm A\sin \Omega t,A\cos \Omega t),  \label{2}
\end{equation}%
where $\Omega $ is the frequency of the light and $A=E_{0}/\Omega $ with $%
E_{0}$ being the amplitude of the electric field $\mathbf{E}(t)=\partial 
\mathbf{A}(t)/\partial t$. The gauge potential satisfies time periodicity, $%
A(t+T)=A(t)$ with $T=2\pi /\Omega $. The plus/minus sign refers to
right/left circular polarization. As long as the photon energy is much
larger than the kinetic energy of the electrons, $\hat{H}^{\eta ,s}(t)$ can
be reduced to an effective \emph{static} (time-independent) Hamiltonian $%
\hat{H}_{\mathrm{eff}}^{\eta ,s}$ using Floquet theory \cite{22,23}. This
method gives results in excellent agreement with experiments \cite{21,23,24}%
. The effective Hamiltonian $\hat{H}_{\mathrm{eff}}^{\eta ,s}$ is defined
through the time evolution operator over one period%
\begin{equation}
\hat{U}(T)=\hat{\mathcal{T}}\exp\left[-i\int_{0}^{T}\hat{H}^{\eta ,s}(t)dt%
\right]=\exp [-i\hat{H}_{\mathrm{eff}}^{\eta ,s}T],  \label{3}
\end{equation}%
where $\hat{\mathcal{T}}$ is the time ordering operator. Using perturbation theory and expanding $\hat{U}(T)$ in the limit of large frequencies $\Omega$, we obtain
\begin{equation}
\hat{H}_{\mathrm{eff}}^{\eta ,s}=\hat{H}_{0}^{\eta ,s}+\frac{1}{\hslash
\Omega }\{[\hat{H}_{+}^{\eta ,s},\hat{H}_{-}^{\eta ,s}]+[\hat{H}_{0}^{\eta
,s},\hat{H}_{+}^{\eta ,s}]-[\hat{H}_{0}^{\eta ,s},\hat{H}_{-}^{\eta
,s}]\}+O(\Omega^{-2}),  \label{4}
\end{equation}%
where $\hat{H}_{m}^{\eta ,s}=(1/T)\int_{0}^{T}e^{-im\Omega t}\hat{H}^{\eta
,s}(t)dt$ is the $m$-th Fourier harmonic of the time-periodic Hamiltonian.
Notice that Eq. (\ref{4}) is only valid under the off-resonance condition $\hslash \Omega \gg t_{j}$ ($%
t_{j}=v\hslash /a$ is the hopping parameter between two nearest neighbors
with $a$ being the lattice constant) \cite{22,new4}. Indeed, when $\Omega\sim t_j$, multiple photon absorption/emission processes must be accounted for, which implies that higher orders in the expansion of $\hat{U}(T)$ should be retained. On the other hand, the condition $\hslash \Omega \gg t_{j}$ implies that the frequency must be much larger than the band width of the system, which is difficult to reach in practice. In fact, for such large frequencies, high energy bands might also contribute to the optical processes. In the present work, we focus on the impact of off-resonant light on low energy bands only and assume that any direct optical processes involving high energy bands only weakly affect the low energy band structure. Still, due to the presence of these high energy processes, the effective power of the incident off-resonant light A is reduced.

Applying on Eq. (1), Eq.\ (4) yields%
\begin{equation}
\hat{H}_{\mathrm{eff}}^{\eta ,s}=v(\eta \hat{\sigma}_{x}\hat{p}_{x}+\hat{%
\sigma}_{y}\hat{p}_{y})+(\Delta\pm\eta \Delta _{\Omega })\hat{\sigma}%
_{z}-s\eta \lambda \hat{\sigma}_{z}+s\eta \lambda ,  \label{5}
\end{equation}%
where $\Delta _{\Omega }=e^{2}v^{2}\hslash ^{2}A^{2}/\hslash ^{3}\Omega ^{3}$
is the effective energy term describing the effects of the circularly
polarized light, which essentially renormalizes the mass of the Dirac
fermions. The plus/minus sign refers to right/left circular polarization.
For right circular polarization the gap is increased in the $K$ valley and
reduced in the $K'$ valley, whereas for left circular
polarization the effect is opposite. After diagonalization we obtain the
eigenvalues%
\begin{equation}
E_{\zeta }^{\eta ,s}=s\eta \lambda +\zeta \sqrt{(v\hslash k)^{2}+(\Delta
+\eta \Delta _{\Omega }-s\eta \lambda )^{2}}  \label{6}
\end{equation}%
and the corresponding eigenfunctions%
\begin{equation}
\Psi _{\zeta }^{\eta ,s}=\frac{e^{ik_{x}x+ik_{y}y}}{\sqrt{L_{x}L_{y}}}\left( 
\begin{array}{c}
\cos \gamma _{k}^{\eta ,s,\zeta }e^{-i\eta \varphi _{k}} \\ 
\sin \gamma _{k}^{\eta ,s,\zeta }%
\end{array}%
\right) .  \label{7}
\end{equation}%
Here $\zeta =\pm 1$ represents the conduction and valence bands,
respectively, $\varphi _{k}=\tan ^{-1}(k_{y}/k_{x})$ with $k=\sqrt{%
k_{x}^{2}+k_{y}^{2}}$, and $\tan \gamma _{k}^{\eta ,s,\zeta }=(E_{\zeta
}^{\eta ,s}-\Delta -\eta \Delta _{\Omega })/v\hbar k$. The impact of
off-resonant light on the band structure is illustrated in Figs.\ 1 and 2
for MoS$_{2}$ ($2\Delta=1.66$ eV, $\lambda =0.0375$ eV, $v=0.5\times10^{5}$
m/s, $a=3.193$ \AA , and $t_{j}=1.10$ eV \cite{5}). We
set $\hslash \Omega=10t_{j}$, which corresponds to a gap variation of
$\Delta_{\Omega }=0.73$ eV for $evA=2.83$ eV. Such a large value of $\hslash \Omega$ ensures that the low energy bands are only affected by virtual emission/absorption processes, while higher energy processes are assumed to only affect the effective power of the incident light (see also Refs.\ \cite{22,new1,new4}). The energy correction $\Delta _{\Omega }
$ can be tuned by varying the amplitude of the electric field or frequency of the light.

Two aspects are worth noticing. First, as mentioned above, the effect of the
off-resonant light is to enhance the gap for the $K$ valley and reduce it
for the $K^{\prime }$ valley ($\Delta _{K}= 3.2$ eV and $\Delta _{K'}= 0.06$
eV in our example). In this case only \emph{one} valley (here
$\eta =-1$) becomes relevant for electronics purposes, enabling almost 100\%
valley polarization of the transport. Second, the SOC-induced splitting is
enhanced in the conduction band, while it remains essentially unchanged in
the valence band. Indeed, in the absence of off-resonant
light the SOC is only active in the valence bands and leaves the conduction
bands almost unaffected. The reduction of the band gap in the $K'$
valley correspondingly empowers SOC-induced splitting in the conduction
bands which become spin-polarized. Nevertheless, we stress that due to
the fact that the two valence bands ($s=\pm 1$) are non-degenerate at $k= 0$
it is still possible to obtain \emph{fully} spin-polarized hole transport by
tuning the Fermi level, while the conduction band can be at
most \emph{partially} spin-polarized. Since the system is fully valley-polarized,
only the conduction band of the $K'$ valley ($\eta =-1$) contributes
to the transport properties discussed in this work.

We calculate the longitudinal
conductivity using the Kubo formula and perform a perturbative expansion in
terms of the short-range impurity potential within the first Born
approximation. We use the Streda \cite{26} version of the Kubo formula \cite%
{27}%
\begin{equation}
\sigma _{i,j}^{-,s}=-\frac{e^{2}\hslash }{4\pi }\int dE\frac{\partial f}{%
\partial E}\text{Tr}[\hat{v}_{i}(\hat{G}^{R}-\hat{G}^{A})\hat{v}_{j}\hat{G}%
^{A}-\hat{v}_{i}\hat{G}^{R}\hat{v}_{j}(\hat{G}^{R}-\hat{G}^{A})],  \label{8}
\end{equation}%
where the velocity components $\hat{v}_{i}$ ($i=x,y$) are given by $\hat{v}%
_{x}=\eta \hat{\sigma}_{x}$ and $\hat{v}_{y}=\hat{\sigma}_{y}$. The
superscript minus represents the $K^{\prime }$ valley, $f$ is the Fermi
distribution function, and $\hat{G}^{A/R}$ are the advanced and retarded
perturbed Green's functions. The unperturbed retarded Green's function is
given by $\hat{G}_{0}^{R}=[E-\hat{H}+i0]^{-1}$. Using Eq.\ (5) with $\eta =-1
$, we obtain 
\begin{equation}
\hat{G}_{0}^{R}=\frac{1}{2}\left[ 
\begin{array}{c}
1+\frac{1}{E_{+}^{-,s}+s\lambda }[(\Delta -\Delta _{\Omega }+s\lambda )\hat{%
\sigma}_{z}-v\hslash (\hat{\sigma}_{x}k_{x}-\hat{\sigma}_{y}k_{y})]%
\end{array}%
\right] \frac{1}{E-E_{+}^{-,s}+i0}.  \label{9}
\end{equation}%
The perturbed Green's function is given by $\hat{G}^{R}=1/[(\hat{G}%
_{0}^{R})^{-1}-\hat{\Sigma}_{R}]$, where $\hat{\Sigma}_{R}$ is the retarded
self-energy. Considering short range randomly distributed impurities in the
first Born approximation, we have 
\begin{equation}
\hat{\Sigma}_{R}=NV_{0}^{2}\int \frac{kdkd\phi }{(2\pi )^{2}}\hat{G}%
_{0}^{R}\approx -iNV_{0}^{2}E_{F}/2\hslash ^{2}v^{2}\text{,}  \label{10}
\end{equation}%
where $N$ is the impurity concentration and $V_{0}$ is the impurity
potential. We obtain 
\begin{equation}
\hat{G}^{R/A}=\frac{1}{2}\left[ 
\begin{array}{c}
1+\frac{1}{E_{+}^{-,s}+s\lambda }[(\Delta -\Delta _{\Omega }+s\lambda )\hat{%
\sigma}_{z}-v\hslash (\hat{\sigma}_{x}k_{x}-\hat{\sigma}_{y}k_{y})]%
\end{array}%
\right] \frac{1}{E\pm i\Gamma -E_{+}^{-,s}},  \label{11}
\end{equation}%
where $\Gamma =-\mathrm{Im}\hat{\Sigma}_{R}$ is the energy broadening due to
the finite quasi-particle lifetime $\tau $. Using Eq.\ (11) in Eq.\ (8) and
addressing the limit of zero temperature with chemical potential $E_{F}$, we
arrive at%
\begin{equation}
\sigma _{xx}^{-,s}=\frac{e^{2}}{2h}\frac{\tau (E_{F}+s\lambda )}{\hslash }%
\left\{ 1-\frac{(\Delta -\Delta _{\Omega }+s\lambda )^{2}}{(E_{F}+s\lambda
)^{2}}\right\} .  \label{12}
\end{equation}%
As expected, the conductivity is enhanced under off-resonant light, since
the effective band gap is reduced. More interestingly, the gap reduction is
accompanied by a spin-polarization of the longitudinal electron flow.
Indeed, since the mass of the carriers is reduced, the impact of SOC is
stronger, leading to a polarization 
\begin{equation}
P=(\sigma _{xx}^{\uparrow }-\sigma _{xx}^{\downarrow })/(\sigma
_{xx}^{\uparrow }+\sigma _{xx}^{\downarrow })=\lambda \frac{E_{F}-(\Delta
-\Delta _{\Omega })}{(E_{F}-(\Delta -\Delta _{\Omega }))E_{F}-2\lambda ^{2}}%
\underset{\Delta _{\Omega }\rightarrow \Delta }{\longrightarrow }\frac{%
\lambda E_{F}}{E_{F}^{2}-2\lambda ^{2}}.  \label{13}
\end{equation}%

The intrinsic Hall conductivity due to
anomalous trajectories of free electrons under the action of the electric
field is expressed in terms of the Berry curvature in $k$-space as \cite%
{5,27,28} 
\begin{equation}
\sigma _{xy}^{\eta,s}=\frac{e^{2}}{\hslash }\int \frac{d^{2}\mathbf{k}}{%
(2\pi )^{2}}[f_{+}(E)-f_{-}(E)]\Omega _{z}^{\eta ,s}(\mathbf{k}),  \label{14}
\end{equation}%
where $f_{\pm }$ denotes the Fermi distributions of the electrons and holes,
respectively. From Eqs.\ (6) and (7), we obtain the Berry curvature as 
\begin{equation}
\Omega _{z}^{\eta ,s}(\mathbf{k})=i\mathbf{z}\cdot \left\langle {\nabla }_{%
\mathbf{k}}\Psi _{\zeta}^{\eta ,s}\left\vert \times \right\vert {\nabla }_{%
\mathbf{k}}\Psi _{\zeta}^{\eta ,s}\right\rangle=\frac{-\eta \hslash
^{2}v^{2}(\Delta+\eta \Delta _{\Omega }-\eta s\lambda )}{2\{(v\hslash
k)^{2}+(\Delta+\eta \Delta _{\Omega }-\eta s\lambda )^{2}\}^{3/2}}.
\label{15}
\end{equation}
Using Eq.\ (\ref{15}) in Eq.\ (\ref{14}) and performing the integral over 
\textbf{k}, we obtain the intrinsic Hall conductivity when
the Fermi level is in the band gap (indicated by a subscript 0)
\begin{equation}
\sigma _{xy,0}^{\eta,s}=\eta \frac{e^{2}}{2h}\text{sgn(}\left\vert \Delta
+\eta \Delta _{\Omega }\right\vert -\eta s\lambda),  \label{16}
\end{equation}%
which yields the quantum spin and valley Hall effects. 
For $\left\vert \Delta -\Delta _{\Omega }\right\vert
>\lambda $ we have $\sigma _{xy,0}^{\eta,s}=\eta e^{2}/2h$, which
results in a vanishing quantum spin Hall effect and finite quantum valley
Hall effect $\sigma _{xy,0}^{v}=\sigma_{xy,0}^{+,s}-\sigma _{xy,0}^{-,s}\sim-e^{2}/h$.
For $\left\vert \Delta-\Delta _{\Omega }\right\vert <\lambda $ we have
$\sigma _{xy,0}^{\eta,s}=-se^{2}/2h$, which results in a vanishing quantum valley
Hall effect and finite quantum spin
Hall effect $\sigma _{xy,0}^{s}=\sigma _{xy,0}^{\eta,+}-\sigma_{xy,0}^{\eta,-}\sim
-e^{2}/h$. Using $\lambda =37.5$ meV we obtain for $\left\vert \Delta
-\Delta _{\Omega }\right\vert =10$ meV a value of $\Delta _{\Omega }=0.82$
eV for $evA=1.65$ eV, which may be varied by alteration of $\Delta _{\Omega }
$ via the intensity of the off-resonant light and is consistent with Fig.\
2. These results can be compared with Ref. \cite{6} where the quantum spin
Hall effect is zero in the limit of \textit{zero off-resonant light}, while
the quantum valley Hall effect is similar to what we obtain above in Eq.\
(16), see Eq.\ (12d) of Ref.\ \cite{6}. Note that the SOC is stronger in WS$%
_{2}$ (107.5 meV) than in MoS$_{2}$ (37.5 meV) so that the quantum spin Hall
effect is easier to detect \cite{5}.

When the Fermi level is in the conduction band (indicated by a subscript 1),
the Hall conductivity is
\begin{equation}
\sigma _{xy,1}^{\eta,s}=\eta\frac{e^{2}}{2h} \frac{\Delta + \eta\Delta _{\Omega
}-\eta s\lambda }{E_{F}-\eta s\lambda}.  \label{17}
\end{equation}
This anomalous conductivity is similar to that reported in Refs.\ 
\cite{5,6} in the limit $\Delta _{\Omega }\rightarrow 0$. However, the
valley selectivity introduced by the circularly polarized off-resonant light
dramatically changes the situation. Indeed, as mentioned in Refs.\ \cite{5,6},
for monolayer MoS$_{2}$ in the \textit{absence of off-resonant light} both
valleys contribute equally to the spin and valley Hall conductivities.
However, in the \textit{presence of off-resonant light}, since the effective
band gap becomes valley-dependent, the spin Hall effect is dominated by one
valley only ($K'$ in the present case) so that it is enhanced,
while the valley Hall effect is correspondingly reduced. Since
the system is fully valley-polarized, only the $K'$ conduction band contributes,
so that the spin ($\sigma _{xy}^{s}=\sigma _{xy,1}^{-,+}-\sigma
_{xy,1}^{-,-}$) and valley ($\sigma _{xy}^{v}=\sigma
_{xy,1}^{-,+}+\sigma _{xy,1}^{-,-}$) Hall conductivities are%
\begin{equation}
\sigma _{xy}^{s}=-\frac{e^{2}}{h}\left( \frac{\lambda (-\Delta +\Delta
_{\Omega }+E_{F})}{E_{F}^{2}-\lambda ^{2}}\right),\quad\sigma _{xy}^{v}=-%
\frac{e^{2}}{h}\left( \frac{(\Delta -\Delta _{\Omega })E_{F}-\lambda ^{2}}{%
E_{F}^{2}-\lambda ^{2}}\right) .  \label{18}
\end{equation}%
When the gap is quenched ($\Delta \rightarrow \Delta _{\Omega }$) the spin
Hall effect is simply proportional to the longitudinal polarization, see
Eq.\ (\ref{13}). The spin and valley Hall conductivies obtained from Eq.\
(18) can be compared to their counterparts $\sigma _{xy}^{s,0}$ and $\sigma
_{xy}^{v,0}$ in the absence of light (see, e.g., Eq.\ (12e) in Ref.\ \cite{6}%
) and we obtain 
\begin{equation}
\sigma _{xy}^{s}=\left( 1-\frac{\Delta _{\Omega }}{\Delta -E_{F}}\right) 
\frac{\sigma _{xy}^{s,0}}{2},\quad\sigma _{xy}^{v}=-\left( 1-\frac{\Delta
_{\Omega }E_{F}}{\Delta E_{F}-\lambda ^{2}}\right) \frac{\sigma _{xy}^{v,0}}{%
2},  \label{19}
\end{equation}%
which explicitly reveals the role of the light-induced gap.
The extrinsic corrections to the anomalous velocity from side jump and skew
scattering have been explicitly calculated in the case of a gapped two-dimensional Dirac
Hamiltonian in Ref.\ \cite{27}. Interestingly, they vanish when the chemical potential
approaches the gap. By its large gap (1.66 eV), this is the case for MoS$_{2}$.
Additionally, skew scattering contributions are inversely proportional
to the impurity concentration and thus vanish in the limit of strong impurity
scattering, compare Ref.\ \cite{new3}.

The last aspect to be discussed is the
enhancement of the orbital magnetic moment. Indeed, in classical
electromagnetism, charges moving with a velocity $v$ along a loop of
diameter $D$ generate an \emph{orbital} magnetic moment $\mu _{\mathrm{orb}%
}=Dev/4$. Recently, values up to $\mu _{\mathrm{orb}}\approx 26$
$\mu _{\mathrm{B}}$ have been reported in 5 nm wide carbon nanotubes \cite{29}. In
general, while the orbital contribution to the magnetization is vanishingly
small in 3$d$ transition metal ferromagnets, it turns out to become
significant in systems involving orbital degrees of freedom such as
nanotubes and graphene-like structures \cite{28,30}. The orbital magnetic
moment can be related to the Berry curvature through the relation \cite%
{28,30} 
\begin{equation}
\mu _{\mathrm{orb}}^{-,s}(k)=\frac{e}{\hslash }E_{+}^{-,s}(k)\Omega
_{z}^{-,s}(k)=\frac{e}{\hslash }E_{+}^{-,s}(k)\frac{\hslash ^{2}v^{2}(\Delta
-\Delta _{\Omega }+s\lambda )}{2\{(v\hslash k)^{2}+(\Delta -\Delta _{\Omega
}+s\lambda )^{2}\}^{3/2}}.
\end{equation}%
For finite $\Delta $ or $\Delta _{\Omega }$ the orbital magnetic moment has
a peak at $k=0$. For zero SOC we obtain for $|\Delta -\Delta _{\Omega }| = 30$
meV a single valley orbital magnetic moment of 35 Bohr magnetons. This may
be varied by alteration of $\Delta _{\Omega }$ by modifying the intensity of
the off-resonant light. The orbital magnetic moment turns out to be
inversely proportional to the band gap.

The corresponding orbital magnetization is \cite{28,30} 
\begin{equation}
M_{\mathrm{orb}}^{-,s}=\frac{e}{\hslash }\int \frac{d^{2}k}{(2\pi )^{2}}%
\left(\mu _{\mathrm{orb}}^{-,s}(k)+\frac{e}{\hslash}[E_{F}-E_{+}^{-,s}(k)]%
\Omega _{z}^{-,s}(k)\right)  \label{21}
\end{equation}%
and we calculate analytically%
\begin{equation}
M_{\mathrm{orb}}^{-,s}=\frac{eE_{F}}{2h}\left( 1-\frac{\Delta -\Delta
_{\Omega }+s\lambda }{E_{F}+s\lambda }\right).  \label{22}
\end{equation}%
Interestingly, this expression has a similar structure as the Hall
conductivity, Eq.\ (\ref{17}), and can be enhanced by reducing the band gap
using off-resonant light. As a reference, for a Fermi energy of 100 meV and
a layer thickness of typically 0.6 nm, we would have an orbital
magnetization of 0.05 Tesla, which is easily detectable and tunable by
varying $\Delta _{\Omega }$. The orbital magnetization can be probed by
various experimental techniques, including susceptibility measurements,
electron paramagnetic resonance, x-ray magnetic circular dichroism, and
neutron diffraction \cite{31,32,33}. The orbital contribution to the
magnetization affects a variety of properties and phenomena such as the
nuclear magnetic resonance \cite{34} and electron paramagnetic resonance 
\cite{35} g-tensors, which both are related to the derivative of the orbital
magnetization as well as to the magnetic susceptibility, the orbital
magnetoelectric response \cite{36,37,38}, and the quantum spin Hall
conductivity \cite{39}.

We propose to use off-resonant circularly polarized
light to enable valley-polarized nanoelectronics in group-VI dichalcogenide
monolayers such as MoS$_{2}$. We theoretically demonstrate that under such
illumination the band gaps of the $K$ and $K^{\prime }$ valleys are
oppositely tuned, leading to 100\% valley polarization. This phenomenon
leads to a number of remarkable effects: (i) Enhancement of the longitudinal
conductivity, accompanied by an increase in the
spin-polarization of the flowing electrons, (ii) enhancement of the
intrinsic spin Hall effect, together with a reduction of
the intrinsic valley Hall effect, and (iii) enhancement of
the orbital magnetic moment and orbital magnetization. Our predictions
cna be realized experimentally by the setup used in Ref.\ \cite{new5} for studying WS$_2$.
The discussed findings expand the horizon of fundamental investigations of the
electronic properties of two-dimensional dichalcogenide systems and presents
promising opportunities to the design of tunable phototransistors \cite%
{15,16,17,18,19,20}, photothermoelectric devices \cite{40}, and related
transport devices.

\begin{acknowledgments}
A.\ M.\ acknowledges fruitful discussions with Dr.\ T.\ Korn.
\end{acknowledgments}

\begin{figure}[ht]
\begin{center}
\includegraphics[width=0.7\columnwidth,clip]{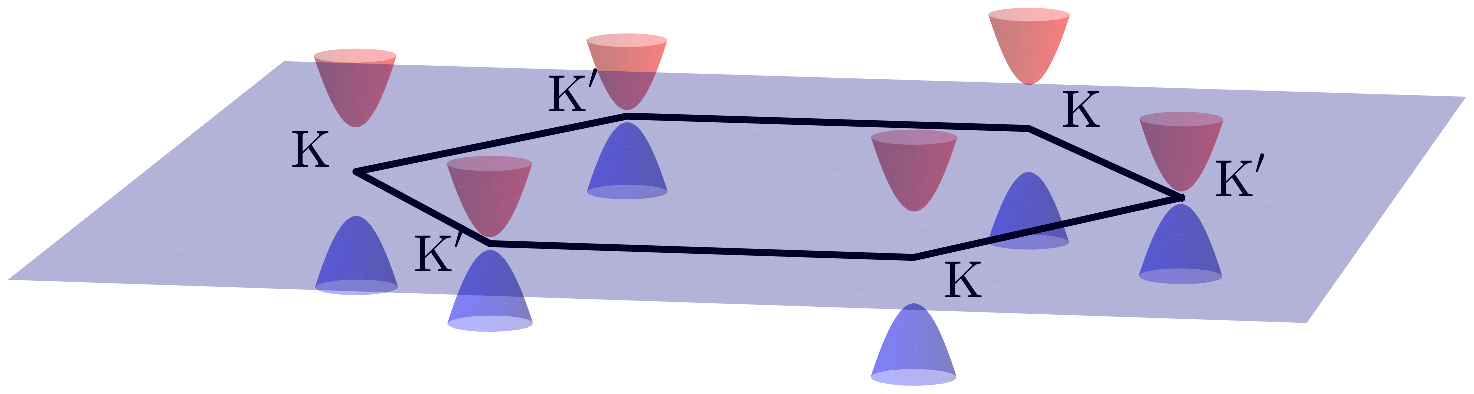}
\end{center}
\caption{Brillouin zone of monolayer MoS$_{2}$ and schematic electronic
structure in the presence of off-resonant light and absence of intrinsic SOC.
}
\label{fig:1}
\end{figure}

\begin{figure}[ht]
\begin{center}
\includegraphics[width=0.5\columnwidth,clip]{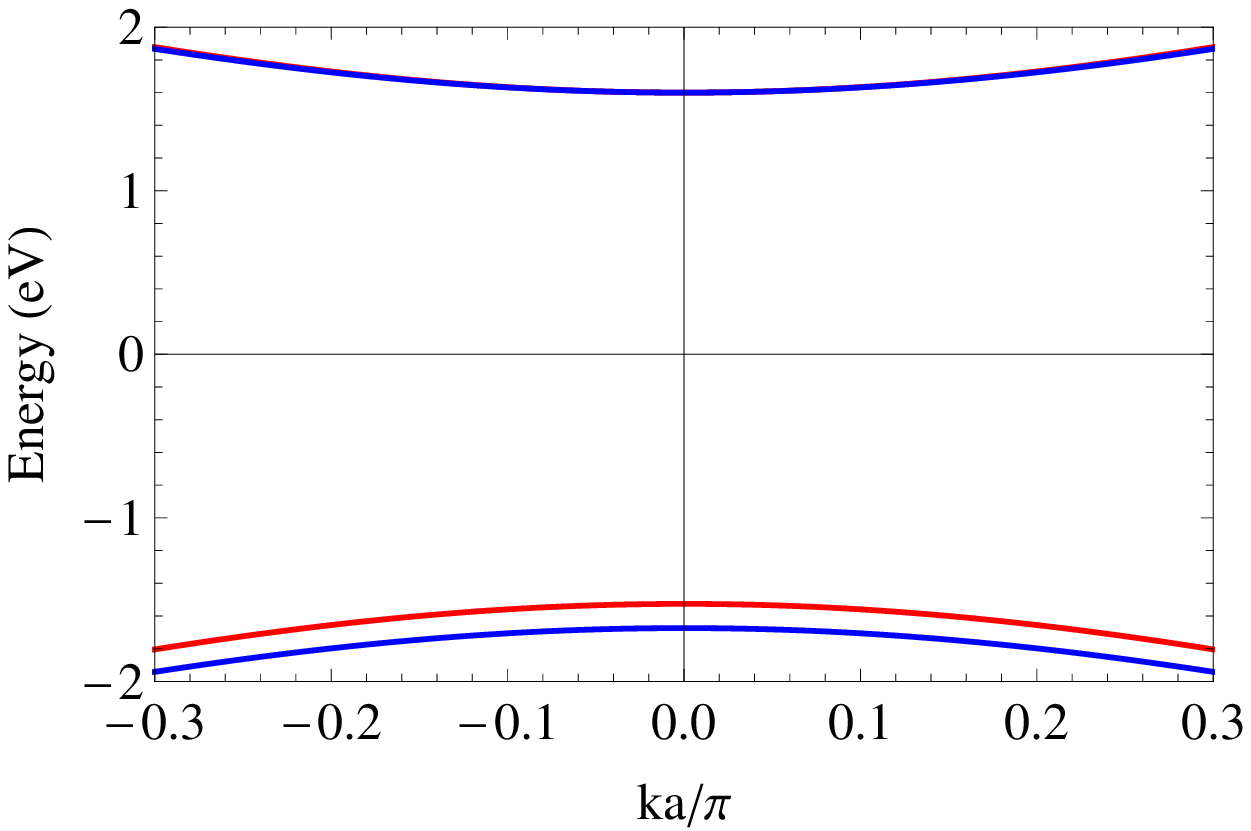} %
\includegraphics[width=0.5\columnwidth,clip]{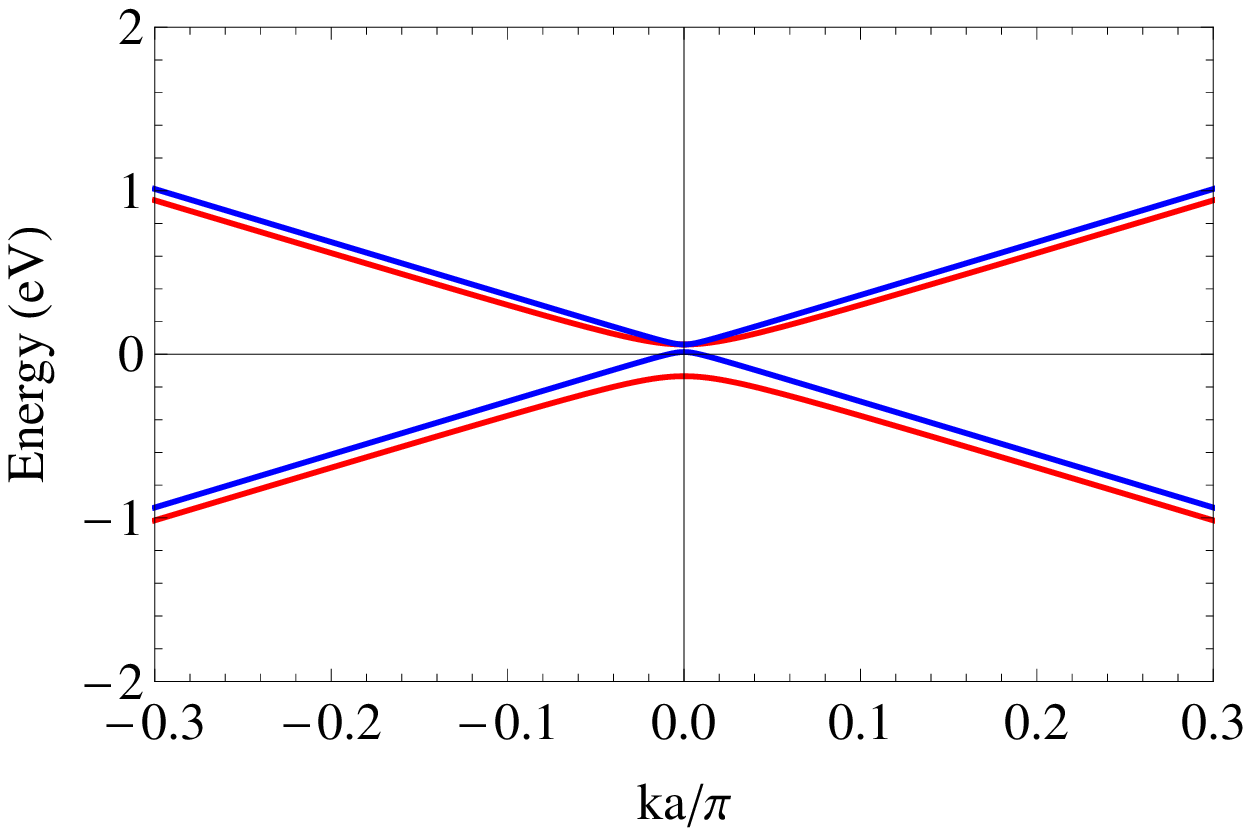}
\end{center}
\caption{Band structure of monolayer MoS$_{2}$ in the presence of both
off-resonant light and intrinsic SOC for the (top) $K$ and (bottom) $%
K^{\prime }$ valley.}
\label{fig:2}
\end{figure}


\begin{thebibliography}{99}
\bibitem{1} A. K. Geim and I. V. Grigorieva, Nature 499, 419 (2013).

\bibitem{2} B. Radisavljevic, A. Radenovic, J. Brivio, V. Giacometti, and A.
Kis, Nat. Nanotechnol. 6, 147 (2011).

\bibitem{3} K. S. Novoselov, A. K. Geim, S. Morozov, D. Jiang, Y. Zhang, S.
Dubonos, I. Grigorieva, and A. A. Firsov, Science 306, 666 (2004).

\bibitem{4} A. H. Castro Neto, F. Guinea, N. M. R. Peres, K. S. Novoselov,
and A. K. Geim, Rev. Mod. Phys. 81, 109 (2009).

\bibitem{5} D. Xiao, G.-B. Liu, W. Feng, X. Xu, and W. Yao, Phys. Rev. Lett.
108, 196802 (2012).

\bibitem{6} Z. Li and J. P. Carbotte, Phys. Rev. B 86, 205425 (2012).

\bibitem{7} K. F. Mak, C. Lee, J. Hone, J. Shan, and T. F. Heinz, Phys. Rev.
Lett. 105, 136805 (2010).

\bibitem{8} A. Splendiani, L. Sun, Y. Zhang, T. Li, J. Kim, C. Y. Chim, G.
Galli, and F. Wang, Nano Lett. 10, 1271 (2010).

\bibitem{9} A. Kuc, N. Zibouche, and T. Heinz, Phys. Rev. B 83, 245213
(2011).

\bibitem{10} T. Cao, G. Wang, W. Han, H. Ye, C. Zhu, J. Shi, Q. Niu, P. Tan,
E. Wang, B. Liu, and J. Feng, Nat. Commun. 3, 887 (2012).

\bibitem{11} K. F. Mak, K. He, J. Shan, and T. F. Heinz, Nat. Nanotechnol.
7, 494 (2012).

\bibitem{12} H. Zeng, J. Dai, W. Yao, D. Xiao, and X. Cui, Nat. Nanotechnol.
7, 490 (2012).

\bibitem{13} S.Wu, J. S. Ross,G.-B. Liu, G. Aivazian, A. Jones, Z. Fei, W.
Zhu, D. Xiao, W. Yao, D. Cobden, and X. Xu, Nat. Phys. 9, 149 (2013).

\bibitem{14} A. M. Jones, H. Yu, N. J. Ghimire, S. Wu, G. Aivazian, J. S.
Ross, B. Zhao, J. Yan, D. G. Mandrus, D. Xiao, W. Yao, and X. Xu, Nat.
Nanotechnol. 8, 634 (2013).

\bibitem{15} Z. Yin, H. Li, L. Jiang, Y. Shi, Y. Sun, G. Lu, Q. Zhang, X.
Chen, H. Zhang, ACS Nano 6, 74 (2012).

\bibitem{16} H. S. Lee, S. W. Min, Y. G. Chang, M. K. Park, T. Nam, H. Kim,
J. H. Kim, S. Ryu, and S. Im, Nano Lett. 12, 3695 (2012).

\bibitem{17} W. Choi, M. Y. Cho, A. Konar, J. H. Lee, G. -B. Cha, S. C.
Hong, S. Kim, J. Kim, D. Jena, J. Joo, S. Kim, Adv. Mater. 24, 5832 (2012).

\bibitem{18} M. Fontana, T. Deppe, A. K. Boyd, M. Rinzan, A. Y. Liu, M.
Paranjape, and P. Barbara, Sci. Rep. 3, 1634 (2013).

\bibitem{19} M. Shanmugam, C. A. Durcan, B. Yu, Nanoscale 4, 7399 (2012).

\bibitem{20} O. L. Sanchez, D. Lembke, M. Kayci, A. Radenovic, and A. Kis,
Nat. Nanotechnol. 8, 497 (2013).

\bibitem{21} Y. H. Wang, H. Steinberg, P. J. Herrero, and N. Gedik, Science
342, 453 (2013).

\bibitem{22} T. Kitagawa, T. Oka, A. Brataas, L. Fu, E. Demler, Phys. Rev. B
84, 235108 (2011).

\bibitem{23} T. Kitagawa, M. A. Broome, A. Fedrizzi, M. S. Rudner, E. Berg,
I. Kassal, A. A. Guzik, E. Demler, and A. G. White, Nat. Commun. 3, 882
(2012).

\bibitem{24} M. C. Rechtsman, J. M. Zeuner, Y. Plotnik, Y. Lumer, D.
Podolsky, F. Dreisow, S. Nolte, M. Segev, and A. Szameit, Nature 496, 196
(2013).

\bibitem{new2} K. F. Mak, C. Lee, J. Hone, J. Shan, and T. F. Heinz, Phys.
Rev. Lett. 105, 136805 (2010).

\bibitem{new1} M. Ezawa, Phys. Rev. Lett. 110, 026603 (2013).

\bibitem{new4} \'A. G\'omez-Le\'on, P. Deplace, and G. Platero, Phys. Rev. B 89, 205408 (2014).

\bibitem{25} J. S. Barriga, A. Varykhalov, J. Braun, S.-Y. Xu, N. Alidoust,
O. Kornilov, J. Minar, K. Hummer, G. Springholz, G. Bauer, R. Schumann, L.
V. Yashina, H. Ebert, M. Z. Hasan, and O. Rader, Phys. Rev. X 4, 011046
(2014).

\bibitem{26} P. Streda, J. Phys. C 15, L717 (1982).

\bibitem{27} N. A. Sinitsyn, J. E. Hill, H. Min, J. Sinova, and A. H.
MacDonald, Phys. Rev. Lett. 97, 106804 (2006).

\bibitem{28} D. Xiao, W. Yao, and Q. Niu, Phys. Rev. Lett. 99, 236809 (2007).

\bibitem{new3} S. Onoda, N. Sugimoto, and N. Nagaosa, Phys. Rev. Lett. 97,
126602 (2006).

\bibitem{29} E. D. Minot, Y. Yaish, V. Sazonova, and P. L. McEuen, Nature
428, 536 (2004).

\bibitem{30} T. Thonhauser, Int. J. Mod. Phys. B 25, 1429 (2011).

\bibitem{31} L. L. Hirsh, Rev. Mod. Phys. 69, 607 (1997).

\bibitem{32} R. M. White, Quantum Theory of Magnetism (Springer, Berlin,
2007).

\bibitem{33} Magnetism and Synchrotron Radiation, edited by E. Beaurepaire,
H. Bulou, F. Scheurer, and J.-P. Kappler, Springer Proceedings in Physics
Vol. 133 (Springer, Berlin, 2010). See also the references therein.

\bibitem{34} T. Thonhauser, D. Ceresoli, A. A. Mostofi, N. Marzari, R.
Resta, and D. Vanderbilt, J. Chem. Phys. 131, 101101 (2009).

\bibitem{35} D. Ceresoli, U. Gerstmann, A. P. Seitsonen, and F. Mauri, Phys.
Rev. B 81, 060409(R) (2010).

\bibitem{36} A. M. Essin, J. E. Moore, and D. Vanderbilt, Phys. Rev. Lett.
102, 146805 (2009).

\bibitem{37} A. Malashevich, I. Souza, S. Coh, and D. Vanderbilt, New J.
Phys. 12, 053032 (2010).

\bibitem{38} A. M. Essin, A. M. Turner, J. E. Moore, and D. Vanderbilt,
Phys. Rev. B 81, 205104 (2010).

\bibitem{39} S. Murakami, Phys. Rev. Lett. 97, 236805 (2006).

\bibitem{new5} E. J. Sie, J. W. McIver, Y.-H. Lee, L. Fu, J. Kong, and N. Gedik,
ArXiv:1407.1825 (2014).

\bibitem{40} M. Buscema, M. Barkelid, V. Zwiller, H. S. J. van der Zant, G.
A. Steele, and A. C. Gomez, Nano Lett. 13, 358 (2013).

\end{thebibliography}
\end{document}